\documentclass[epj]{webofc}
\usepackage[varg]{txfonts}
\usepackage{amsmath}
\usepackage{slashed}
\usepackage{bbold}
\usepackage{hyperref}

\wocname{EPJ Web of Conferences}

\woctitle{CONF12}

\begin{document}

\selectlanguage{english}
\title{Three-point vertex functions in Yang-Mills Theory and QCD\\ in Landau gauge}

\author{Adrian L. Blum\inst{1}\fnsep\thanks{\email{adrian.blum @ uni-graz.at}} \and
        Reinhard Alkofer\inst{1}\fnsep\thanks{\email{reinhard.alkofer @ uni-graz.at}}\and
        Markus Q. Huber\inst{1}\fnsep\thanks{\email{markus.huber @ uni-graz.at}} \and
				Andreas Windisch\inst{2}\fnsep\thanks{\email{windisch@physics.wustl.edu}}
}

\institute{Institute of Physics, University of Graz, NAWI Graz, Universit\"atsplatz 5, 8010
Graz, Austria 
\and
           Department of Physics, Washington University, St. Louis, MO, 63130, USA 
}

\abstract{
Solutions for the three-gluon and quark-gluon vertices from Dyson-Schwinger equations and the three-particle irreducible formalism are discussed. Dynamical quarks (``unquenching'') change the three-gluon vertex via the quark-triangle diagrams which themselves include fully dressed quark-gluon vertex functions. On the other hand, the quark-swordfish diagram is, at least with the model used for the two-quark-two-gluon vertex employed here, of minor importance. For the leading tensor structure of the three-gluon vertex the ``unquenching'' effect can be summarized for the nonperturbative part as a shift of the related dressing function towards the infrared.
}

\maketitle

\section{Introduction}
\label{intro}

All the information about a quantum field theory is contained in its $n$-point functions. E.g., once the $n$-point functions of quantum chromodynamics (QCD) are computed, they can be used in a next step to calculate hadron properties. On the level of three-point functions, especially the three-gluon and quark-gluon vertices are of interest and they have thus been the objectives of many non-perturbative studies in recent years. For instance, in the Landau gauge they have been studied using lattice \cite{Skullerud:2003qu,Lin:2005zd,Kizilersu:2006et,Cucchieri:2008qm,Maas:2011se,Oliveira:2016muq,Athenodorou:2016oyh,Duarte:2016ieu,Sternbeck:2016tgv} and continuum methods \cite{Davydychev:2000rt,LlanesEstrada:2004jz,Binger:2006sj,Alkofer:2008tt,Alkofer:2008dt,Binosi:2011wi,Hopfer:2013np,Aguilar:2013ac,Pelaez:2013cpa,Ahmadiniaz:2012xp,Windisch:2012de,Huber:2012kd,Aguilar:2013vaa,Blum:2014gna,Eichmann:2014xya,Rojas:2013tza,Windisch:2014th,Gracey:2014mpa, Hopfer:2014th,Williams:2014iea,Aguilar:2014lha,Pelaez:2015tba,Mitter:2014wpa,Williams:2015cvx,Cyrol:2016tym,Ahmadiniaz:2016qwn,Binosi:2016wcx,Aguilar:2016lbe}. For the Coulomb gauge recent results can be found in \cite{Huber:2014isa,Vastag:2015qjd,Campagnari:2016wlt,Campagnari:2016chq}.

The three-gluon vertex captures the property of self-interactions between the gauge bosons in non-Abelian theories and thus is supposed to be linked to confinement. In view of this, the by now compelling evidence that at least in Landau gauge the coefficient function of the leading tensor structure changes sign for small Euclidean momenta is puzzling: Approaching from the ultraviolet at the momenta which are relevant for hadron physics the strength of the three-gluon interaction is {\em decreasing} contrary to the expectation of a strong confining-type force. In the far infrared the three-gluon interaction changes from an anti-screening to a screening-type interaction. However, at these momenta, firstly, the gluon propagator is dominated by the ghost contributions (i.e., by gauge-fixing), and secondly, hadronic properties have decoupled from the Yang-Mills degrees of freedom. Certainly, this truly unexpected scenario deserves further studies. 

In such an approach, hadron properties cannot be understood without further knowledge about the quark-gluon vertex which is the crucial quantity for the coupling of the Yang-Mills to the matter sector. In recent years it has become evident that this vertex displays an astonishing complexity, especially for momenta around a few hundred MeV. As these are the momenta directly relevant for the properties of hadrons, the phenomenological success of an approach to hadron physics via QCD Green functions requires a detailed understanding of the emerging structures in the quark-gluon vertex.

In the following we report on our studies undertaken with the objective to understand the interplay between the three-gluon and the quark-gluon vertices. 
We present and discuss solutions for these vertices from Dyson-Schwinger equations (DSEs) and the three-particle irreducible (3PI) formalism with a special emphasis of the back-coupling of quarks (``unquenching'') on the three-gluon vertex.

\section{The unquenched three-gluon vertex}

\subsection{Input}

\begin{figure}[b]
\centering
\includegraphics[width=135mm,clip]{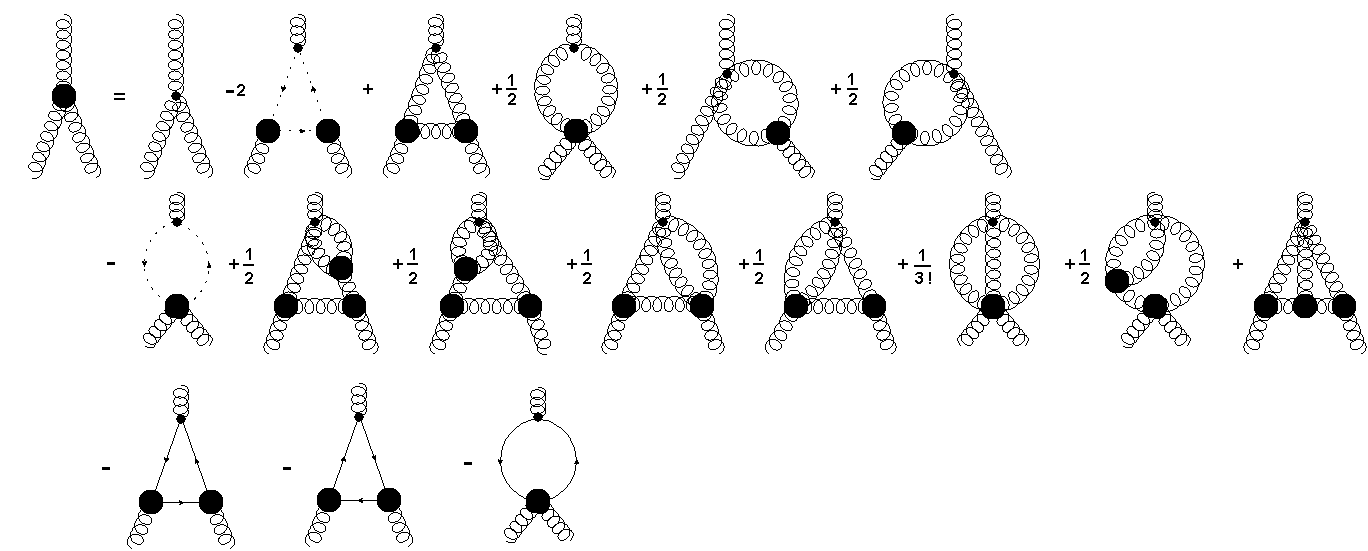}
\caption{The full DSE of the (unquenched) three-gluon vertex. Here and in all other figures, internal propagators are dressed, and thick blobs denote dressed vertices, wiggly lines gluons, dashed ones ghosts, and continuous ones quarks. This figure and Figs.~\ref{truncatedthreeglvertDSE}~-~\ref{quarkpropagator} were created with \textit{JaxoDraw} \cite{Binosi:2003yf}.}
\label{fullthreeglvertDSE}
\end{figure}

The full DSE for the three-gluon vertex is shown in Fig.\ref{fullthreeglvertDSE}. The first two lines describe the Yang-Mills part of the three-gluon vertex, while (direct) unquenching effects are solely due to the three diagrams in the last line. Here, the first two diagrams in this line are called quark-triangle diagrams and the third one quark-swordfish diagram.

This equation for the three-gluon vertex is coupled to higher $n$-point functions which makes for the purpose of a numerical solution the introduction of a truncation inevitable. A widely-used truncation scheme consists in neglecting two-loop diagrams (part of the second line) and diagrams with vertices that do not have a tree-level counterpart (e.g., the first diagram in the second line), see, however, below. This truncation proved to be quite successful for the three-point functions in the Yang-Mills sector, for a detailed discussion, see, e.g., refs.~\cite{Huber:2012kd,Blum:2014gna,Eichmann:2014xya,Huber:2016tvc,Cyrol:2016tym}. Furry's theorem applies to the quark-triangle diagrams. Thus, the color symmetric parts cancel each other while the color anti-symmetric parts add up so that only one quark-triangle diagram needs to be calculated by taking a factor of two into account \cite{Blum:2015lsa}.

\begin{figure}[t]
\centering
\includegraphics[width=135mm,clip]{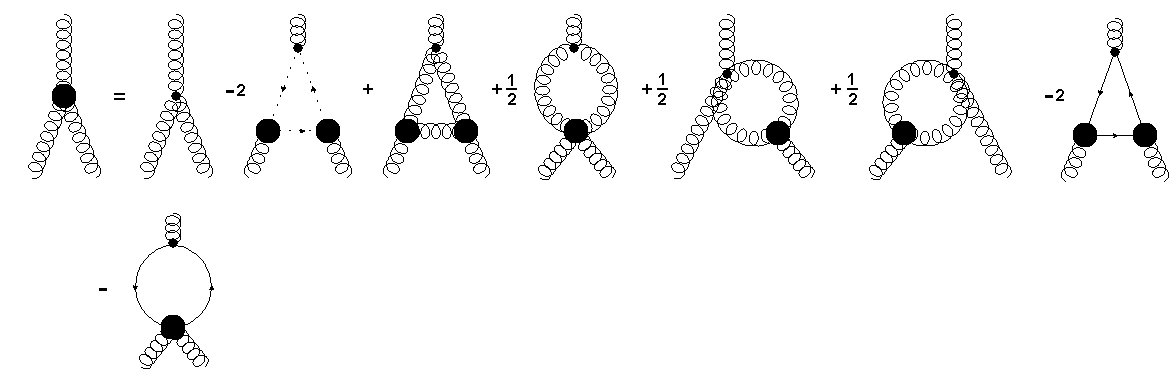}
\caption{The truncated DSE of the (unquenched) three-gluon vertex.}
\label{truncatedthreeglvertDSE}
\end{figure}

In this work we deviate from the truncation scheme introduced above and include also the quark-swordfish diagram which contains the two-quark-two-gluon vertex, a non-primitively divergent correlation function.
For the two-quark-two-gluon vertex a model will be employed which will be described in detail in section \ref{2q2gl-vertex}. The truncated DSE for the three-gluon vertex used for numerical calculations therefore has the form depicted in Fig. \ref{truncatedthreeglvertDSE}. 

Next, an appropriate basis for the three-gluon vertex has to be found. This task is facilitated by the fact that in Landau gauge the full dynamics of the theory is described by the transverse degrees of freedom only \cite{Fischer:2008uz}. The three-gluon vertex can be decomposed into ten longitudinal and four transverse basis tensors where consequently only the transverse tensors have to be considered. Furthermore, it has been shown that the dominant contribution stems from the tree-level tensor structure and that the other three transverse tensors are negligible \cite{Eichmann:2014xya,Sternbeck:2016tgv}. We will therefore focus on the dominant tree-level tensor which leads to the following parametrization for the three-gluon vertex (suppressing the color factor $f^{abc}$):
\begin{align}
 \Gamma^{A^{3}}_{\mu\nu\rho}(p,q,k=-p-q) = D^{A^{3}}(p^2,q^2,\cos(\alpha))\;\Gamma^{A^{3},(0)}_{\mu\nu\rho}(p,q,k=-p-q)\,.
\end{align}
Here $D^{A^{3}}(p^2,q^2,\cos(\alpha))$ denotes the dressing function, $\Gamma^{A^{3},(0)}_{\mu\nu\rho}(p,q,k=-p-q)$ is the tree-level tensor, and $\alpha$ is the angle between the momenta $p$ and $q$.

\begin{figure}[b]
\includegraphics[height=10mm,clip]{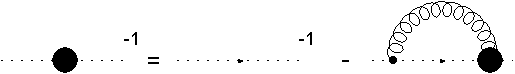}
\\
\vskip3mm
\includegraphics[height=13mm,clip]{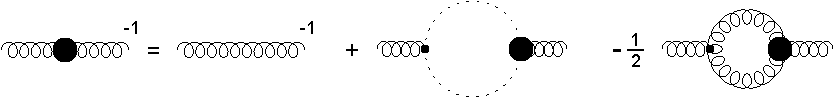}
\caption{The coupled DSEs of the ghost and gluon propagator}
\label{ghost and gluon DSE}
\end{figure} 

Furthermore, to calculate the three-gluon vertex the ghost and gluon propagators,
\begin{align}
D^{G}(p^2) =-\frac{G(p^2)}{p^2}\,, \qquad
D_{\mu\nu}(p^2) = \left(\delta_{\mu\nu}-\frac{p_{\mu}p_\nu}{p^2}\right)\frac{Z(p^2)}{p^2}\,,
\end{align}
are needed. To this end, the coupled and truncated system of both Yang-Mills propagators in Fig. \ref{ghost and gluon DSE} is solved. In doing so, as an intermediate step, a model for the three-gluon vertex in the gluon propagator DSE is employed which is tuned in such a way that two-loop contributions are effectively included~\cite{Huber:2012kd}.

For the ghost-gluon vertex that appears in the ghost-triangle diagram (the second diagram in 
Fig.~\ref{truncatedthreeglvertDSE}) the bare vertex is used. This is a good approximation since it has been found in earlier works that the ghost-gluon vertex shows only in the midmomentum regime a minor deviation from the bare vertex \cite{Schleifenbaum:2004id, Huber:2012kd}. The influence on the three-gluon vertex is only small \cite{Blum:2014gna}. On the other hand, the four-gluon vertex (the fourth diagram in Fig. \ref{truncatedthreeglvertDSE}) is described by a model \cite{Blum:2014gna,Eichmann:2014xya} which captures the qualitative aspects of so far performed calculations of the four-gluon vertex \cite{Cyrol:2014kca,Binosi:2014kka}.

\subsection{The quark-gluon vertex}

\begin{figure}[b]
\includegraphics[width=135mm,clip]{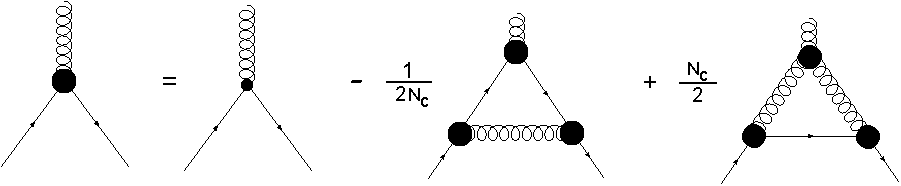}
\caption{The quark-gluon vertex in the 3PI formalism.}
\label{qgv}
\end{figure}

The connection between the Yang-Mills and the matter sector in the three-gluon vertex is provided by the quark-triangle and the quark-swordfish diagrams. The former depends on the dressed quark-gluon vertex. We will consider the quark-gluon vertex in the 3PI formalism which implies that all vertices are dressed, see Fig.~\ref{qgv}. As for the three-gluon vertex we have to find a parametrization of the quark-gluon vertex in terms of basis tensors. The quark-gluon vertex can in general be decomposed into eight transverse and four longitudinal basis tensors where again in Landau gauge only the transverse part is required:
\begin{align}
\Gamma^{qgv}_{\mu}(p,q,\Delta=p-q) = \sum \limits_{i=1}^{8}g_{i}(p^2,q^2,\cos(\alpha)) b_{\mu}^{(i)}\,.
\end{align}

\begin{figure}[b]
\includegraphics[width=85mm,clip]{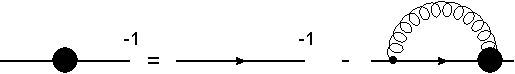}
\caption{The quark propagator DSE.}
\label{quarkpropagator}
\end{figure}

We will use two different basis sets \cite{Windisch:2014th, Hopfer:2014th}. Internally we will use the transversely projected naive basis
\begin{align}
\left\{ \begin{matrix}
     \mathbb{1}\\
      \slashed{p}\\
    \slashed{q}\\
   \slashed{p} \slashed{q}
\end{matrix} \right\} \otimes \left\{ \begin{matrix} \gamma_{\mu}\\ p_{\mu} \\ q_{\mu} \end{matrix} \right \}
    \Rightarrow P_{\mu\nu}(\Delta)\, \left\{ \begin{matrix}
     \mathbb{1}\\
      \slashed{\Sigma}\\
    \slashed{\Delta}\\
   \slashed{\Sigma} \slashed{\Delta}
\end{matrix} \right\} \otimes \left\{\begin{matrix} \gamma_{\nu}\\ \Sigma_{\nu} \\ \Delta_{\nu} \end{matrix} \right \} \,,
\end{align}
where $\Delta$ and $\Sigma$ are $\Delta = p-q$ and $\Sigma = \frac{(p + q)}{2} $ with $p$ and $q$ the anti-quark and quark momenta, respectively. However, it is easier to project out the dressing functions in an orthonormal basis set which will therefore be used externally. The use of two different basis sets implies that one has to transform from one basis to the other within each iteration step.

\begin{figure}[tb]
\includegraphics[width=0.49\textwidth]{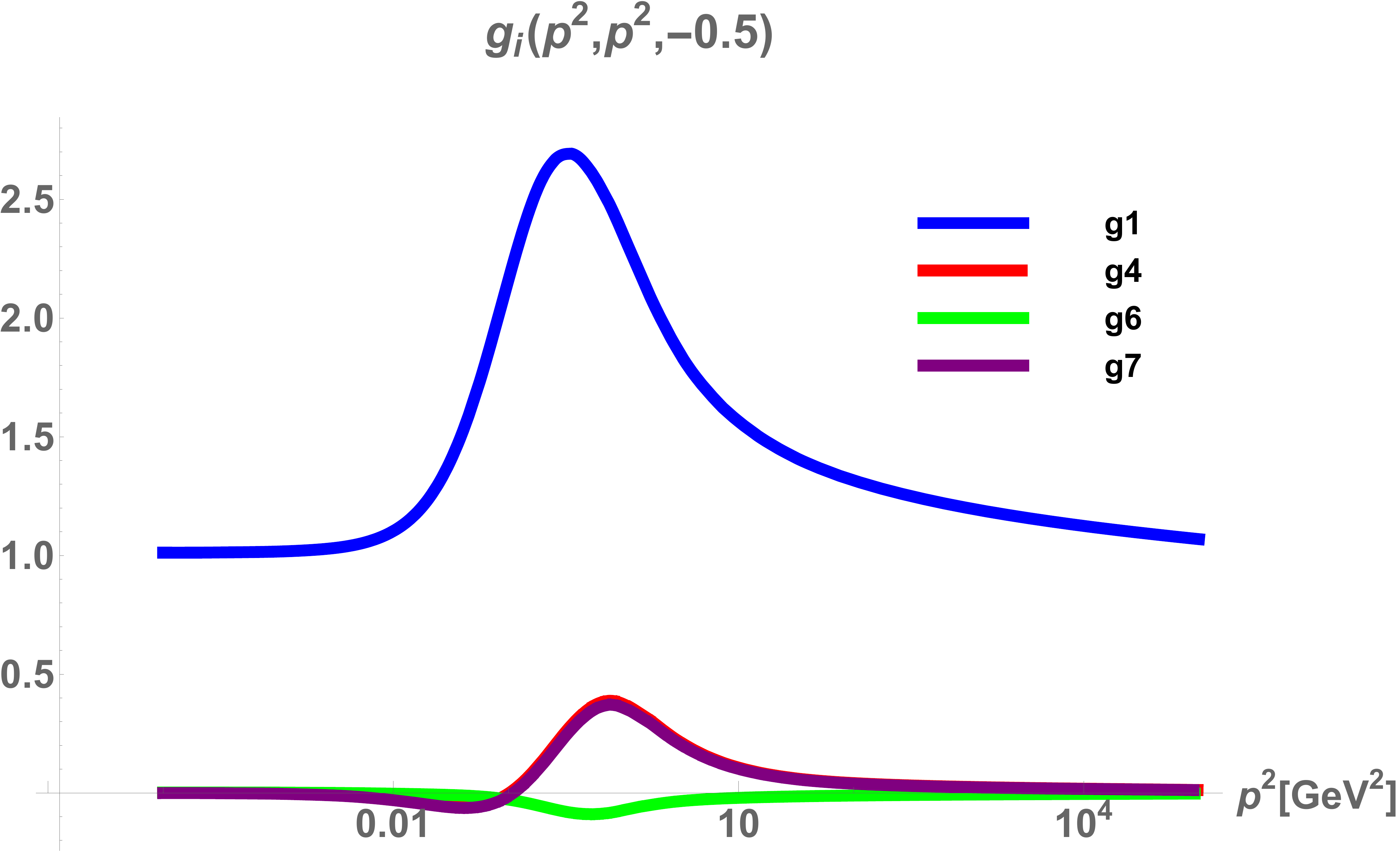}
\hfill
\includegraphics[width=0.49\textwidth]{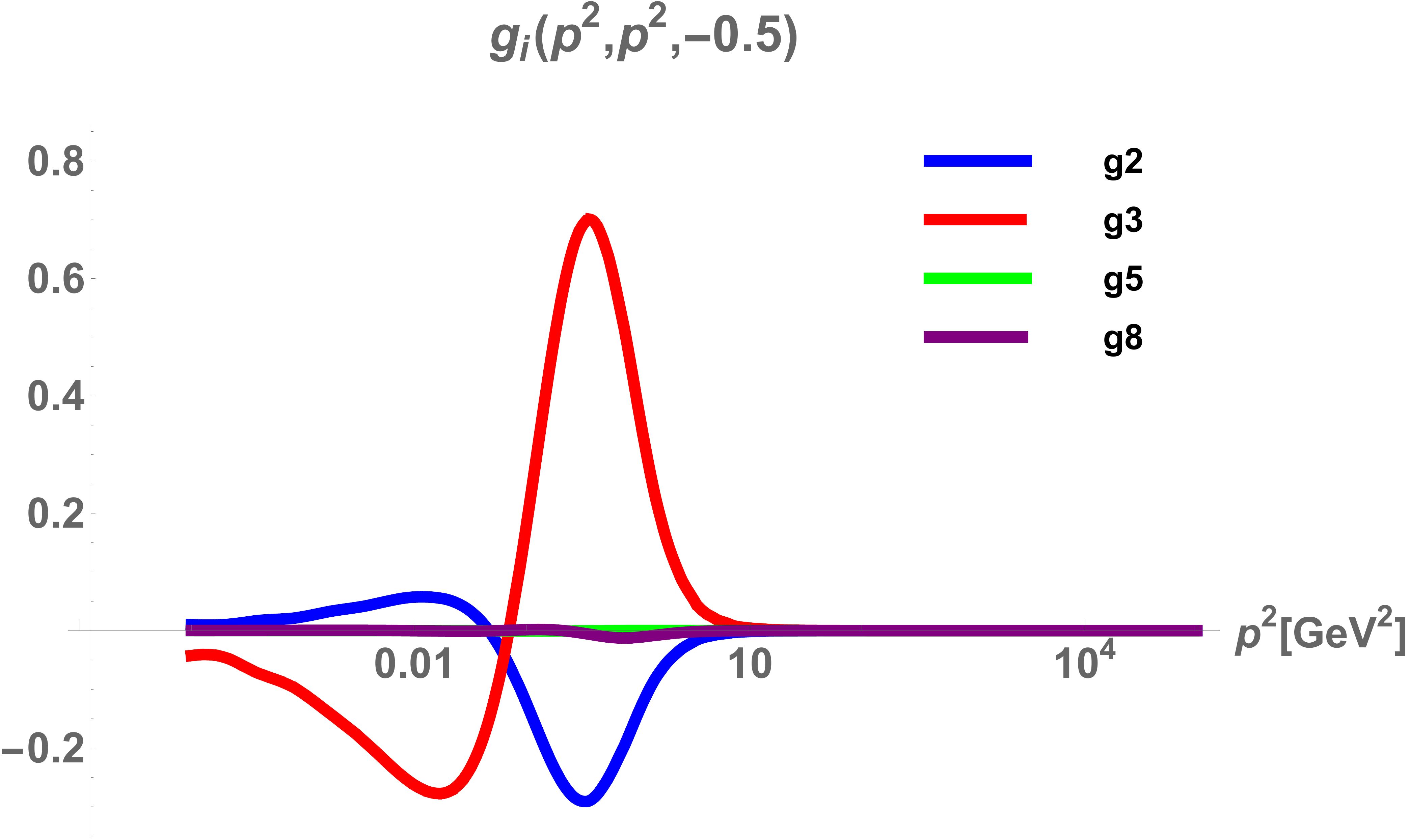}
\caption{The dressing functions of the quark-gluon vertex at the symmetric point corresponding to the eight transverse tensor structures. The left plot shows the results for the chirally symmetric and the right plot for the chirally anti-symmetric structures.}  
\label{fig:qug}
\end{figure}

Coupled to the quark-gluon vertex is the quark propagator 
\begin{align}
S(p)&=\frac{1}{-i\slashed{p}A(p^2)+B(p^2)}
    =Z_{f}(p^2) \frac{i\slashed{p}+M(p^2)}{p^2+M^2(p^2)}\;,
\end{align}		
with the quark wave function renormalization $Z_{f}(p^2)=1/A(p^2)$ and the quark mass function $M(p^2) = B(p^2)/A(p^2)$. Its DSE is depicted in Fig.~\ref{quarkpropagator}.
The coupled system of quark propagator and quark-gluon vertex is then solved \cite{Hopfer:2013np,Windisch:2014th, Hopfer:2014th} and used as input for the unquenched three-gluon vertex.
The results for the quark-gluon vertex dressing functions are shown in Fig.~\ref{fig:qug}.\footnote{The differences to the results shown in \cite{Blum:2015lsa} are due to different parameters for the three-gluon vertex model used as input.}

\subsection{A model for the two-quark-two-gluon vertex}
\label{2q2gl-vertex}

The last diagram in Fig. \ref{truncatedthreeglvertDSE} that needs consideration is the quark-swordfish diagram. As mentioned earlier, this diagram depends on the two-quark-two-gluon vertex which is not primitively divergent. We model the vertex based on an idea that was developed in \cite{Mitter:2014wpa}. Crucial for this is the fact that each of the eight tensor structures of the quark-gluon vertex alone breaks gauge invariance. However, if further terms, called $T_{gauge}$ below, are ``added'' to the Lagrangian, gauge invariance can be restored. 

A simple example mentioned in \cite{Mitter:2014wpa} to illustrate the general idea of this gauge completion is given by the tree-level tensor structure. The tree-level tensor structure appears in the form $\bar{q}\slashed{A}q$ in the Lagrangian which breaks gauge invariance. However, there is a further term $\bar{q}\slashed{\partial}q$ so that both terms together form a gauge invariant quantity namely the covariant derivative $\bar{q}\slashed{D}q$. 

In a similar manner, the restoring terms can be found for the other basis tensors of the quark-gluon vertex. These are, however, more complex than for the tree-level tensor in the sense that they do not only give a contribution at the level of the quark-gluon vertex but also at the level of higher vertices like the two-quark-two-gluon and the two-quark-three-gluon vertices, $T_{gauge} \propto \mathcal{O}(\bar{q}A q) + \mathcal{O}(\bar{q}AAq) + \mathcal{O}(\bar{q}AAAq)\,.$
If one assumes that to each dressing of the quark-gluon vertex ($\mathcal{O}(\bar{q}Aq)$) there is only one dressing function $\lambda_{gauge}$ for the complete $T_{gauge}$, one can deduce the higher-order vertices from the quark-gluon vertex as obtained from its DSE described in the last subsection. We find that the expression of $T_{gauge}$ at the level of $\mathcal{O}(\bar{q}A q)$ is proportional to the tensor $b_\mu^{(4)}+b_\mu^{(7)}$ of the internal basis. In the three-gluon vertex DSE only ($\mathcal{O}(\bar{q}AAq)$) appears. To be concrete, based on \cite{Mitter:2014wpa} our model for the two-quark-two-gluon vertex is :
\begin{align}
\Gamma^{2q2gl,ab}_{\mu\nu}(p,q,k,p-q-k) = -\frac{i}{4}\,g^2\,\lambda^{\bar{q}qAA}(\bar{p}) T_{\mu \nu \rho}  \left( (k+q)_\rho T^a T^b + (k-p)_\rho T^b T^a\right),
\end{align}
where $T_{\mu \nu \rho} = \{[\gamma_\mu, \gamma_\nu], \gamma_\rho\}$ and $\bar{p}$ is the average over all momenta. Taking into account that $g_4\approx g_7$, we obtain for the dressing 
\begin{align} 
\lambda^{\bar{q}qAA}(\bar{p}) = \frac{g_{(4)}(\bar{p}^{2},\bar{p}^{2},-0.5)}{(\sqrt{3/4}\bar{p}^{2})}.       
\end{align}
Here, we have approximated the dressing function of the two-quark-two-gluon vertex by using the average momentum in the dressing function of the quark-gluon vertex in order to resolve the second gluon momentum.

\section{Results}

\begin{figure}[tb]
\includegraphics[width=0.475\textwidth,clip]{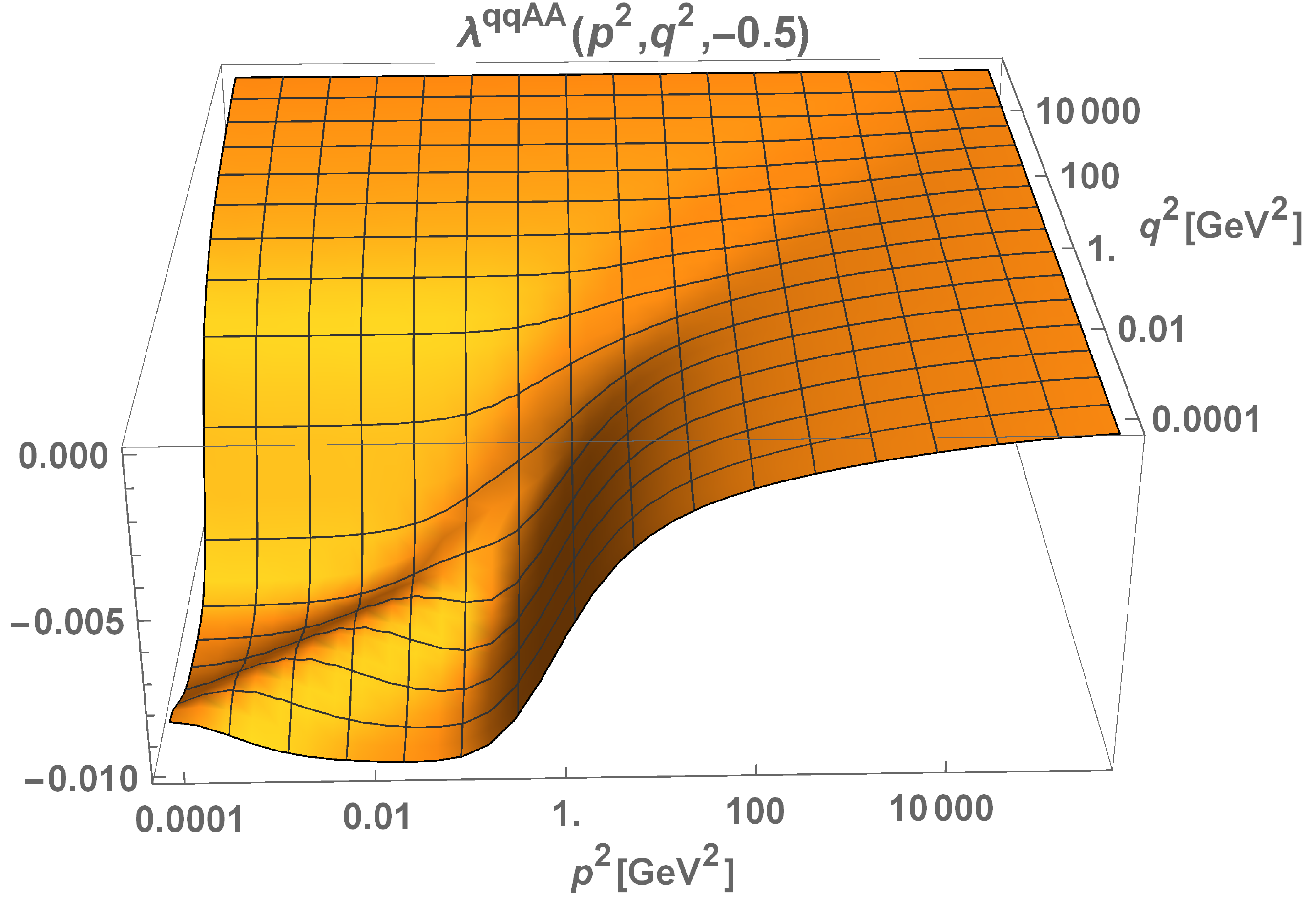}
\hfill
\includegraphics[width=0.475\textwidth,clip]{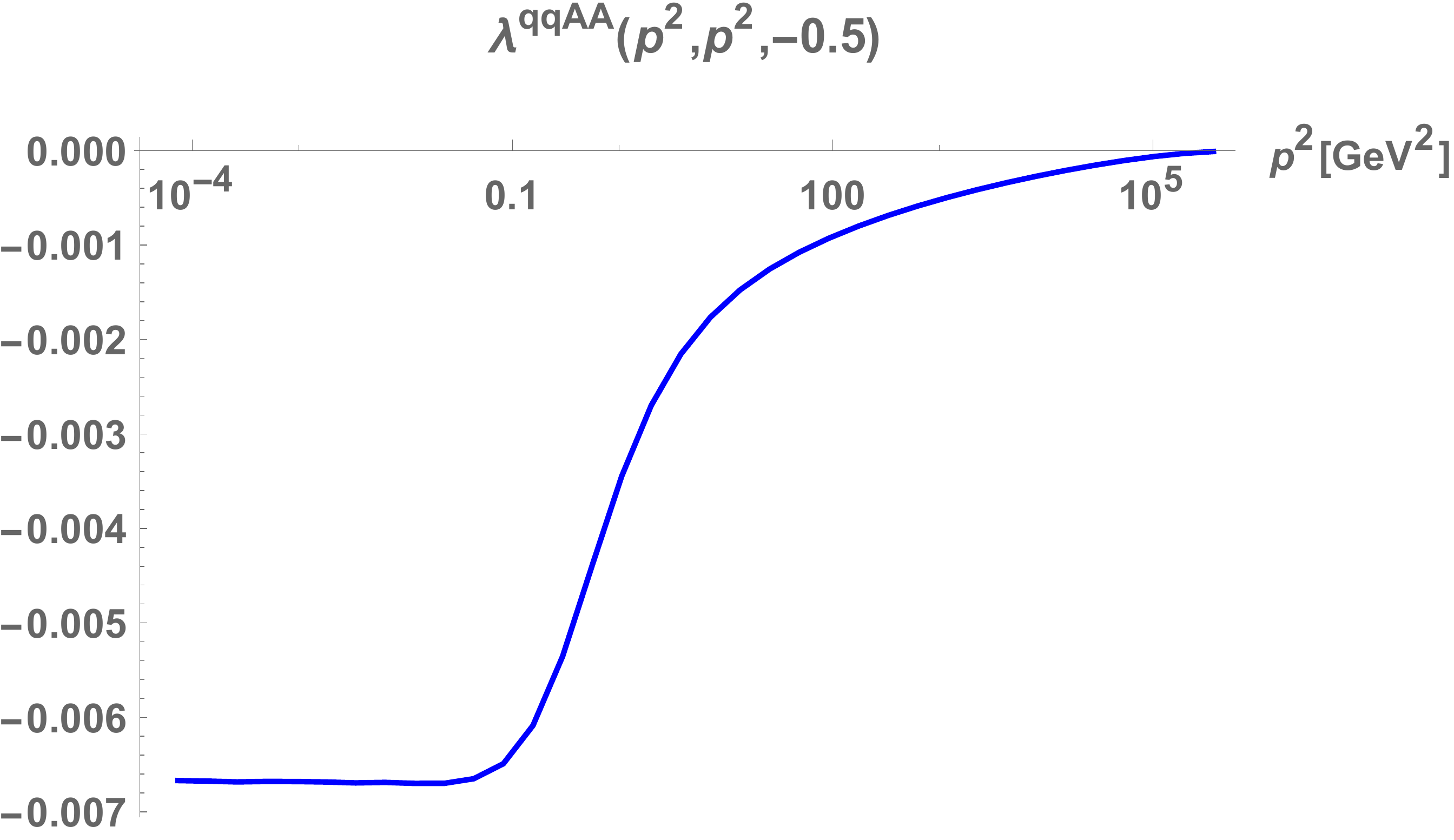}
\caption{The quark-swordfish diagram that appears in the three-gluon vertex DSE. On the left-hand side it is plotted at a fixed angle $\cos(\alpha)=-0.5$ and on the right-hand side it is plotted at the symmetric point, i.e. $q^2=p^2$ and $\cos(\alpha)=-0.5$.}
\label{quark swordfish diagram}
\end{figure}

With the model for the two-quark-two-gluon vertex described in the last subsection one can compute the quark-swordfish diagram independently from the other diagrams in the three-gluon vertex DSE since it is a static diagram and remains therefore unchanged by the iteration process. 
The result for this diagram is presented in Fig. \ref{quark swordfish diagram}.

As the plot reveals, the contribution of the quark-swordfish diagram is very small and therefore it has a minor impact on the three-gluon vertex. However, this is not yet a final conclusion, since the employed model of the two-quark-two-gluon vertex lacks justification through an explicit calculation. 

\begin{figure}[tb]
\includegraphics[width=0.49\textwidth,clip]{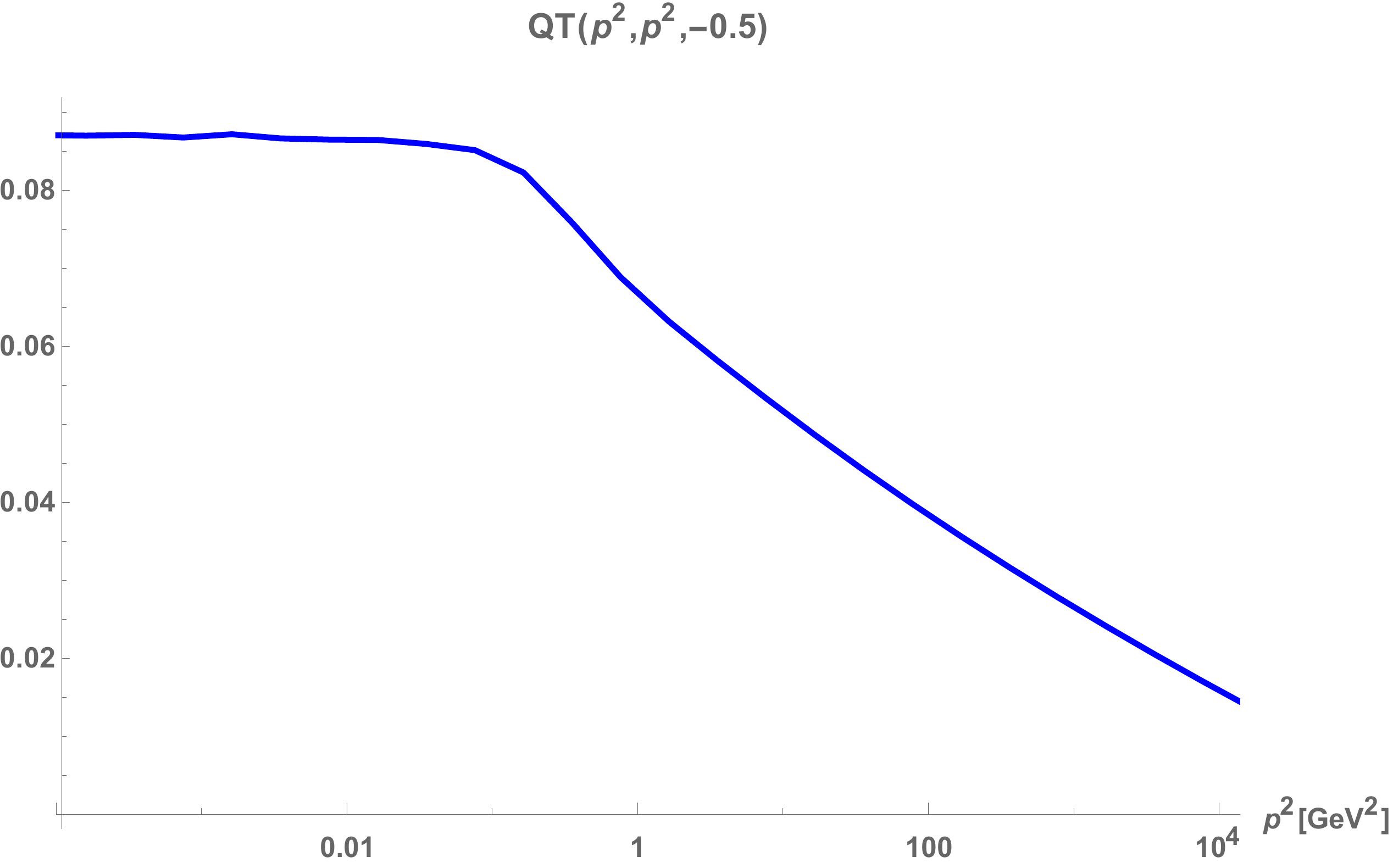}
\hfill
\includegraphics[width=0.49\textwidth,clip]{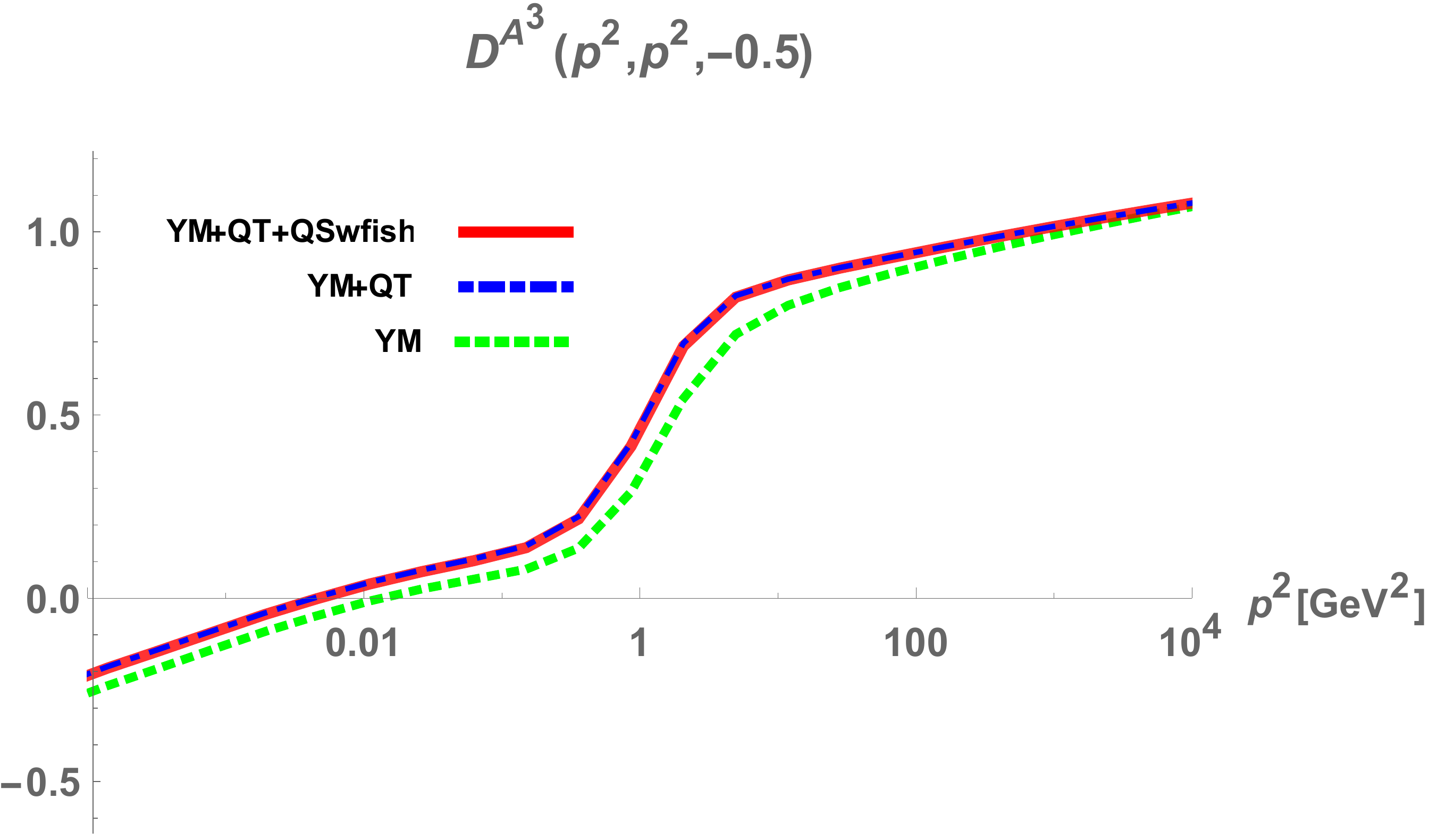}
\caption{\textit{Left:} The contribution from the quark-triangle diagram for symmetric momentum configuration $q^2=p^2$, $p\cdot q = -p^2/2$ calculated with the tree-level component of the quark-gluon vertex.
\textit{Right:} The fully iterated three-gluon vertex. The green line represents a pure Yang-Mills calculation. For the blue line the quark-triangle was added and for the red line the quark-triangle and the quark-swordfish diagrams were added.}
\label{QTb1b1+unquenched3glvert}
\end{figure}

Likewise, the quark-triangle diagram can be computed independently from the iteration process. In doing so, we focus on the tree-level tensor structure of the quark-gluon vertex as it turned out that, despite the not so small magnitude of non-tree-level dressing functions, this yields by far the main contribution. The resulting contribution from the quark-triangle diagram for this setup is depicted in Fig. \ref{QTb1b1+unquenched3glvert}.\footnote{The sign is different to the one shown in \cite{Blum:2015lsa}. This was due to an overall sign error in the kernel. The sign agrees with the results in \cite{Williams:2015cvx}.}

Finally, a full iteration of the three-gluon vertex for $N_f=1$ has been performed using \textit{DoFun} and \textit{CrasyDSE} \cite{Huber:2011qr,Huber:2011xc}. Hereby we proceeded in three steps. At first, only the Yang-Mills part of the three-gluon vertex was iterated (the green curve in Fig.~\ref{QTb1b1+unquenched3glvert}). This calculation was then extended by adding solely the quark-triangle (the blue line) and both the quark-triangle and quark-swordfish diagrams (the red line) to the iteration procedure. The contribution of the quark-swordfish diagram is as expected negligible. In Fig.~\ref{QTb1b1+unquenched3glvert} we can moreover see that the zero-crossing is pushed towards the infrared regime. A further striking effect is that the unquenched three-gluon vertex lies above the quenched vertex which is the opposite of what one finds for the gluon propagator. These results are in agreement with the findings in \cite{Williams:2015cvx}, where the system of quark propagator and all three three-point functions was solved with fixed Yang-Mills propagator input. As is evident from Fig.~\ref{QTb1b1+unquenched3glvert}, the unquenching effect on the leading component of the three-gluon vertex can be summarized as a global shift of the nonperturbative part towards the infrared.

It should be noted that within this truncation the mechanism of changing the sign of the dressing function is extremely robust, since the relevant negative contribution stems from the ghost triangle which is logarithmically divergent \cite{Huber:2012kd,Pelaez:2013cpa,Aguilar:2013vaa,Blum:2014gna,Eichmann:2014xya}. All contributions from quarks and gluons become constant at low momenta, so finally this dressing function will always become negative. The only possibility to change that rests in the ghost-swordfish diagram, which has not been investigated up to now. Lattice calculations show a clear suppression of the tree-level dressing function at low momenta and are compatible with a zero crossing \cite{Athenodorou:2016oyh,Duarte:2016ieu,Sternbeck:2016tgv}, but the few points with negative values have still somewhat large statistical errors.

\section{Conclusions}

The unquenching effects in the three-gluon vertex arise due to the quark-triangle and the quark-swordfish diagrams in the three-gluon vertex DSE. We presented a fully iterated result for the three-gluon vertex with both diagrams included. 
The quark-gluon vertex in the quark-triangle diagram has been taken from a separate calculation of the coupled quark propagator and quark-gluon vertex DSEs.
However, we restricted ourselves to the dominant tree-level tensor structure in the quark-gluon vertex.

We went beyond standard truncation schemes by including also the quark-swordfish diagram. For its calculation we employed a model for the two-quark-two-gluon vertex based on an effective restoration of gauge invariance and the calculated quark-gluon vertex. We found the contribution of the quark-swordfish diagram to be negligible within our setup. A comparison between the quenched and the unquenched three-gluon vertex shows that unquenching effects lead to a shift of the leading component of the three-gluon vertex towards the infrared and thus to an increased strength at momenta relevant for hadron physics.

\section*{Acknowledgements}
Funding by the FWF (Austrian science fund) through the Doctoral Program ``Hadrons in Vacuum, Nuclei and Stars'', Contract W1203-N16, and under Contract P 27380-N27 is gratefully acknowledged. AW acknowledges support through the FWF Schr\"odinger Fellowship J3800-N27 and the U.S. Department of Energy, Office of Science, Office of Nuclear Physics under Award Number \#DE-FG02-05ER41375.

\bibliography{literature_confXII}

\end{document}